\documentclass[submission,copyright,creativecommons]{eptcs}
\usepackage{proof}
\usepackage{amssymb,amsmath,amsthm}
\usepackage{times}
\usepackage{color}
\usepackage{latexsym}
\usepackage{amssymb}
\usepackage{diagrams}
\usepackage{url}
\usepackage{microtype}

\author{Jos\'e Esp\'\i{}rito Santo\\Centro de Matem\'atica\\Universidade do
Minho\\Portugal\\ \texttt{jes@math.uminho.pt}}
\title{A note on strong normalization in classical natural deduction}

\def\titlerunning{A note on strong normalization in classical natural deduction}
\def\authorrunning{J.~Esp\'\i{}rito~Santo}

\newcommand{\D}{\mathcal{D}}

\newcommand{\lb}{\lambda}
\newcommand{\pair}[2]{\langle#1,#2\rangle}
\newcommand{\proj}[2]{\pi_{#1}(#2)}
\newcommand{\inj}[2]{\mathsf{in}_{#1}(#2)}
\newcommand{\case}[5]{\mathsf{case}(#1,#2.#3,#4.#5)}

\newcommand{\connective}{\bigcirc}

\newcommand{\E}{\mathcal{E}}
\newcommand{\Ei}{\mathcal{E}_{\supset}} %------ E implication
\newcommand{\Ec}{\mathcal{E}_{\wedge}} %------ E conjunction
\newcommand{\Ed}{\mathcal{E}_{\vee}} %------ E disjunction
\newcommand{\Es}{\mathcal{E}_{\connective}} %------ E any

\newcommand{\hole}[1]{[#1]}%-------------------- hole
\newcommand{\holec}{\_}%----------------hole contents
\newcommand{\ehole}{\hole{\holec}}%-------------------- empty hole

\newcommand{\betai}{\beta_{\supset}}%------- beta implication
\newcommand{\betac}{\beta_{\wedge}}%-------- beta conjunction
\newcommand{\betad}{\beta_{\vee}} %-------- beta disjunction
\newcommand{\pii}{\pi_{\supset}}%-------- pi implication
\newcommand{\pic}{\pi_{\wedge}}%-------- pi conjunction
\newcommand{\pid}{\pi_{\vee}}%-------- pi disjunction
\newcommand{\pis}{\pi_{\connective}}%-------- pi any

\newcommand{\Deltad}{\Delta_{\vee}} %-------- Delta disjunction
\newcommand{\rhoi}{\rho_{\supset}}%------- Delta implication
\newcommand{\rhoc}{\rho_{\wedge}}%-------- Delta conjunction
\newcommand{\rhod}{\rho_{\vee}} %-------- Delta disjunction
\newcommand{\rhos}{\rho_{\connective}} %-------- Delta any

\newcommand{\ld}{\lambda_{\Delta}}
\newcommand{\ldmd}{\lambda_{\Delta}^{-\vee}}%--------lambda-Delta minus disjunction
\newcommand{\ldmdc}{\lambda_{\Delta}^{-\vee\wedge}}%--------lambda-Delta minus disjunction and conjunction

\newcommand{\dmg}[1]{#1^{\blacklozenge}}

\newcommand{\mc}[1]{#1^{\dagger}}%--------------minus conjunction

\newtheorem{defn}{Definition}
\newtheorem{lemma}{Lemma}
\newtheorem{cor}{Corollary}
\newtheorem{prop}{Proposition}
\newtheorem{thm}{Theorem}

\begin{document}

\maketitle

% remove in final version!
%\pagestyle{plain}

\begin{abstract}
In the context of natural deduction for propositional classical logic, with classicality given by the inference rule \emph{reductio ad absurdum}, we investigate the De Morgan translation of disjunction in terms of negation and conjunction. Once the translation is extended to proofs, it obtains a reduction of provability to provability in the disjunction-free subsystem. It is natural to ask whether a reduction is also obtained for, say, strong normalization; that is, whether strong normalization for the disjunction-free system implies the same property for the full system, and whether such lifting of the property can be done along the De Morgan translation. Although natural, these questions are neglected by the literature. We spell out the map of reduction steps induced by the De Morgan translation of proofs. But we need to ``optimize'' such a map in order to show that a reduction sequence in the full system from a proof determines, in a length-preserving way, a reduction sequence in the disjunction-free system from the De Morgan translation of the proof. In this sense, the above questions have a positive answer.
\end{abstract}

%\newpage
%\tableofcontents

%-------------------------------------
\section{Introduction}\label{sec:introduction}

In the context of natural deduction for propositional classical logic, with classicality given by the inference rule \emph{reductio ad absurdum} \cite{Prawitz65}, we investigate the De Morgan translation of disjunction in terms of negation and conjunction $A\vee B:=\neg(\neg A\wedge\neg B)$. This translation immediately extends to proof-rules, as recalled in Fig.~\ref{fig:de-Morgan-map-of-proof-rules},
which obtains a reduction of provability to provability in the disjunction-free subsystem.

%------------------------------------------------
\begin{figure}[t]\caption{The de Morgan translation of proofs}\label{fig:de-Morgan-map-of-proof-rules}
$$
\begin{array}{rcccl}
\infer{A_1\vee A_2}{\deduce{A_1}{\vdots}}&\quad&\mapsto&\quad&
\infer[w]{\neg(\neg A_1\wedge\neg A_2)}{\infer{\perp}{\infer{\neg
A_i}{[w:\neg A_1\wedge\neg A_2]}&\deduce{A_i}{\vdots}}}
\end{array}
$$

$$
\begin{array}{rcl}
\infer[x,y]{C}{\deduce{A\vee
B}{\vdots}&\deduce{C}{\deduce{\vdots}{[x:A]}}&\deduce{C}{\deduce{\vdots}{[y:B]}}}&\mapsto&
\infer[k]{C}{\infer{\perp}{\deduce{\neg(\neg A\wedge\neg
B)}{\vdots}&\infer{\neg A\wedge\neg B}{\infer[x]{\neg
A}{\infer{\perp}{[k:\neg
C]&\deduce{C}{\deduce{\vdots}{[x:A]}}}}&\infer[y]{\neg
B}{\infer{\perp}{[k:\neg C]&\deduce{C}{\deduce{\vdots}{[y:B]}}}}}}}
\end{array}
$$
\end{figure}
%------------------------------------------------

It is the case that a reduction is also obtained for, say, strong normalization? To be more precise, is it the case that strong normalization for the disjunction-free system implies the same property for the full-system, and that such lifting of the property can be done along the De Morgan translation, that is, through a mapping of reduction sequences? The answer is not clear at all, nevertheless these questions are neglected by the literature. For instance St\aa lmarck \cite{Stalmarck91} says ``\emph{(...) the strong normalization theorem for the restricted version of first order classical N.D. together with the well-known results on the definability of the rules for $\vee$ and $\exists$ in the restricted system does not imply the normalization theorem for the full system}'', and this author moves on to give a direct proof of strong normalization for the full first order system; but the quoted claim is not supported by technical evidence. On the other hand, a claim in the opposite direction, namely that the normalization theorem for the full system does follow from the normalization for the restricted system, may be thought of as the implicit, unproven assumption behind the approach to classical natural deduction in Prawitz \cite{Prawitz65}, where only the restricted system is studied, after being considered ``\emph{adequate}''.

In this paper we re-examine these questions, for propositional logic. We spell out the map of reduction steps induced by the De Morgan translation of proofs, and observe that the map does not readily lift strong normalization from the disjunction-free system. We need to ``optimize'' such map in order to show that a reduction sequence in the full system from a proof determines, in a length-preserving way, a reduction sequence in the disjunction-free system from the De Morgan translation of the proof. In this sense, strong normalization for the full system does indeed reduce to strong normalization for the disjunction-free system along the De Morgan translation. As we provide a proof of strong normalization for the disjunction-free system, this completes a new proof of strong normalization for the full system.

Our natural deduction system is presented as a variant of the $\ld$-calculus \cite{RehofSorensen94}, and we employ typed $\lb$-terms throughout to code logical derivations. As to reduction rules, in addition to detour conversion and the commuting conversions pertaining to disjunction, one has conversions related to \emph{reductio ad absurdum inferences} - $\rho$-conversions, as we will call them. We do not adopt the conversions by Prawitz for the atomization of the conclusion of r.a.a. inferences \cite{Prawitz65}, but rather the conversions of St\aa lmarck \cite{Stalmarck91} (which go back to Statman \cite{StatmanPhD74}), except that we do not impose any constraint on the $\rho$-conversion for disjunction, in this way following \cite{RehofSorensen94}.

The paper is organized as follows. In Section \ref{sec:background} we present our natural deduction system and its disjunction-free subsystem. In Section \ref{sec:lifting-of-SN} we study the De Morgan translation of proofs, and how it lifts strong normalization from the disjunction-free subsystem to the full system. In Section \ref{sec:SN} we prove strong normalization of the disjunction-free subsystem. %Section \ref{sec:conclusions} concludes.

\section{Background}\label{sec:background}

We present our logical system as a calculus of the $\ld$ family
\cite{RehofSorensen94}. Types/formulas are given by
$$
A,B,C\,::=\,X\,|\,\perp\,|\,A\supset B\,|\,A\wedge B\,|\,A\vee B
$$
We define $\neg A:=A\supset\perp$.

Proof terms:
$$
\begin{array}{rcll}
M,N,P,Q&::=&x&\textrm{(assumption)}\\
&|&\lb x^{A}.M\,|\,MN&\textrm{(implication)}\\
&|&\pair MN\,|\,\proj 1M\,|\,\proj 2M&\textrm{(conjunction)}\\
&|&\inj 1M\,|\,\inj 2N\,|\,\case M{x^A}P{y^B}Q&\textrm{(disjunction)}\\
&|&\Delta k^{\neg A}.M&\textrm{(\emph{reductio ad absurdum})}
\end{array}
$$

\noindent The type annotation in the bound variable of binders will often be
omitted when no confusion arises.

The typing/inference rules are in Fig.~\ref{fig:typing}. $\Gamma$
denotes a set of \emph{declarations} $x:A$ such that a variable is
declared at most one time in $\Gamma$.

%--------------------------------
\begin{figure}[t]\caption{Typing/inference rules}\label{fig:typing}
$$
\begin{array}{c}
\infer[Ass]{\Gamma,x:A\vdash x:A}{}\\ \\
\infer[\supset I]{\Gamma\vdash\lb x^{A}.M:A\supset B}{\Gamma,x:A\vdash B}\qquad
\infer[\supset E]{\Gamma\vdash MN:B}{\Gamma\vdash M:A\supset B&\Gamma\vdash N:A}\\ \\
\infer[\wedge I]{\Gamma\vdash \pair MN:A\wedge B}{\Gamma\vdash
M:A&\Gamma\vdash N:B}\qquad
\infer[\wedge E1]{\Gamma\vdash \pi_1(M):A}{\Gamma\vdash M:A\wedge B}\qquad\infer[\wedge E2]{\Gamma\vdash \pi_2(M):B}{\Gamma\vdash M:A\wedge B}\\ \\
\infer[\vee I1]{\Gamma\vdash\inj 1M:A\vee B}{\Gamma\vdash M:A}\qquad
\infer[\vee I2]{\Gamma\vdash\inj 2N:A\vee B}{\Gamma\vdash N:A}\\ \\
\infer[\vee E]{\Gamma\vdash\case M{x^A}P{y^B}Q:C}{\Gamma\vdash
M:A\vee
B&\Gamma,x:A\vdash P:C&\Gamma,y:B\vdash Q:C}\\ \\
\infer[RAA]{\Gamma\vdash\Delta k^{\neg A}.M:A}{\Gamma, k:\neg A\vdash M:\perp}%\\ \\
%\textrm{(The remaining typing rules are obvious.)}
\end{array}
$$
\end{figure}
%--------------------------------

For the purpose of defining some reduction rules and the translation
of proof terms, it is convenient to arrange the syntax of the system in a
different way:
$$
\begin{array}{rrcl}
\textrm{(Terms)}&M,N,P,Q&::=&V\,|\,\E\hole M\,|\,\Delta k^{\neg A}.M\\
\textrm{(Values)}&V&::=&x\,|\,\lb x.M\,|\,\pair MN\,|\,\inj 1M\,|\,\inj
2N\\
\textrm{(Elim. contexts)}&\E&::=&\ehole N\,|\,\proj
1{\ehole}\,|\,\proj 2{\ehole}\,|\,\case {\ehole}{x^A}P{y^B}Q
\end{array}
$$
\noindent A \emph{value} $V$ ranges over terms representing
assumptions or introduction inferences. $\E$ stands for an
\emph{elimination context}, which is a term representing an
elimination inference, but with a ``hole'' in the position of the
main premiss. $\E\hole M$ denotes the term resulting from filling
the hole of $\E$ with $M$.
%$$
%\E::=\ehole N\,|\,\proj 1{\ehole}\,|\,\proj 2{\ehole}\,|\,\case
%{\ehole}{x^A}P{y^B}Q
%$$
%\noindent $\E\hole M$ denotes the term resulting from filling the
%hole of $\E$ with $M$.

In Fig.~\ref{fig:typing-contexts} one finds the typing rules for
contexts. Without surprise \cite{CurienHerbelinICFP00}, these are
particular cases of sequent calculus inference rules. In a sequent
$\Gamma|A\vdash\E:B$, $A$ is the type of the hole of $\E$ and $B$ is
the type of the term obtained by filling the hole of $\E$ with a
term of type $A$.

%------------------------
\begin{figure}[t]\caption{Typing for contexts}\label{fig:typing-contexts}
$$
\begin{array}{c}
\infer{\Gamma|A\supset B\vdash\ehole N:B}{\Gamma\vdash
N:A}\qquad\infer[(i=1,2)]{\Gamma|A_1\wedge A_2\vdash\proj
i{\ehole}:A_i}{}\\ \\
\infer{\Gamma|A\vee B\vdash\case{\ehole}xPyQ:C}{\Gamma,x:A\vdash
P:C&\Gamma,y:B\vdash Q:C}\\ \\
\infer{\Gamma\vdash\E\hole M:B}{\Gamma\vdash M:A&\Gamma|A\vdash\E:B}
\end{array}
$$
\end{figure}
%----------------------------------

The reduction (or proof transformation) rules are given in
Fig.~\ref{fig:red-rules}. They make use of the following
organization of the definition of elimination contexts: $\E::=\Ei\,|\,\Ec\,|\,\Ed$, where
$$
\Ei::=\ehole N\qquad
\Ec::=\proj 1{\ehole}\,|\,\proj 2{\ehole}\qquad
\Ed::=\case{\ehole}{x^A}P{y^B}Q\\
$$
%$$
%\begin{array}{rcl}
%\Ei&::=&\ehole N\\
%\Ec&::=&\proj 1{\ehole}\,|\,\proj 2{\ehole}\\
%\Ed&::=&\case{\ehole}{x^A}P{y^B}Q\\
%\E&::=&\Ei\,|\,\Ec\,|\,\Ed
%\end{array}
%$$
%--------------------------------
\begin{figure}[t]\caption{Reduction (or proof transformation) rules}\label{fig:red-rules}
\vspace{.25cm} Detour conversion rules:
$$
\begin{array}{rrcll}
(\betai)&(\lb x.M)N&\to&[N/x]M&\\
(\betac)&\proj i{\pair{M_1}{M_2}}&\to&M_i&\textrm{ ($i=1,2$)}\\
(\betad)&\case{\inj iM}{x_1}{P_1}{x_2}{P_2}&\to&[M/x_i]P_i&\textrm{ ($i=1,2$)}\\
\end{array}
$$
Commutative conversion rule:
$$
\begin{array}{rrcl}
(\pis)&\Es\hole{\case MxPyQ}&\to&\case Mx{\Es\hole P}y{\Es\hole
Q}\qquad(\connective=\supset,\wedge,\vee)
\end{array}
$$
\emph{Reductio ad absurdum} rules:
$$
\begin{array}{rrcll}
(\rhos)&\Es\hole{\Delta k.M}&\to&\Delta k'.[\lb z.k'(\Es\hole z)/k]M&(\connective=\supset,\wedge,\vee)\\
(\rhos^{\perp})&\Es^{\perp}\hole{\Delta k.M}&\to&[\lb z.\Es^{\perp}\hole z/k]M&(\connective=\supset,\wedge,\vee\textrm{ and } \Es^{\perp}:\perp)\\
%(\Delta_2)&k'(\Delta k.M)&\to&[k'/k]M&(k':\neg A)\\
(\rho_2)&\Delta k.kM&\to&M&(k\notin M)\\
%(\rho_3)&\Delta k^{\neg\perp}.M&\to&[I/k]M&(I:=\lb x.x)
\end{array}
$$
%$\eta$-rules:
%$$
%\begin{array}{rrcll}
%(\etal)&\lb x.Vx&\to&V&(x\notin V)\\
%(\etad)&\Delta k.kM&\to&M&(k\notin M)
%\end{array}
%$$
\end{figure}
%--------------------------------

For each $R\in\{\beta,\pi,\rho_1,\rho_1^{\perp}\}$, we put $R:=R_{\supset}\cup R_{\wedge}\cup R_{\vee}$. Then we put $\rho:=\rho_1\cup\rho_1^{\perp}\cup\rho_2$.

%We let
%$$
%\begin{array}{rcl}
%\beta&:=&\betai\cup\betac\cup\betad\\
%\pi&:=&\pii\cup\pic\cup\pid\\
%\rho_1&:=&\rhoi\cup\rhoc\cup\rhod\\
%\rho_1^{\perp}&:=&\rhoi^{\perp}\cup\rhoc^{\perp}\cup\rhod^{\perp}\\
%\rho&:=&\rho_1\cup\rho_1^{\perp}\cup\rho_2\\
%\eta&:=&\etad
%\end{array}
%$$
An equivalent definition of $\pi$, $\rho_1$ and $\rho_1^{\perp}$ is
$$
\begin{array}{rrcl}
(\pi)&\E\hole{\case MxPyQ}&\to&\case Mx{\E\hole P}y{\E\hole Q}\\
(\rho_1)&\E\hole{\Delta k.M}&\to&\Delta k'.[\lb z.k'(\E\hole
z)/k]M\\
(\rho_1^{\perp})&\E^{\perp}\hole{\Delta k.M}&\to&[\lb z.\E^{\perp}\hole
z/k]M
\end{array}
$$
%Elimination contexts $\E$ permit elegant definitions.% of $\pi$ and $\rho$.

The reduction rules preserve types (subject reduction property). The case of $\rhoi$ is given in Fig. \ref{fig:SR}, where $W$, $S$ and $(*)$ are applications of weakening, the typing of substitution, and last rule of Fig. \ref{fig:typing-contexts}, respectively,

%------------------------------------------------------
\begin{figure}[t]\caption{Example of type preservation}\label{fig:SR}
$$
\begin{array}{c}
\infer[(*)]{\Gamma\vdash\Ei\hole{\Delta k.M}:B}{\infer[RAA]{\Gamma\vdash\Delta k.M:A\supset B}{\deduce{k:\neg(A\supset B),\Gamma\vdash M:\perp}{\D_1}}&\deduce{\Gamma|A\supset B\vdash \Ei:B}{\D_2}}
\\
\\
\downarrow\\
\\
\infer[RAA]{\Gamma\vdash\Delta k'.[\lb z.k'(\Ei\hole{z})/k]M:B}{\infer[S]{k':\neg B,\Gamma\vdash[\lb z.k'(\Ei\hole{z})/k]M:\perp}{\infer[]{k':\neg B,\Gamma\vdash\lb z.k'(\Ei\hole z):\neg(A\supset B)}{\infer[]{\Delta\vdash k'(\Ei\hole z):\perp}{\infer[Ass]{\Delta\vdash k':\neg B}{}&\infer[(*)]{\Delta\vdash\Ei\hole z:B}{\infer[Ass]{\Delta\vdash z:A\supset B}{}&\infer[W]{\Delta|A\supset B\vdash\Ei:B}{\deduce{\Gamma|A\supset B\vdash\Ei:B}{\D_2}}}}}&\infer[W]{k':\neg B,k:\neg(A\supset B),\Gamma\vdash M:\perp}{\deduce{k:\neg(A\supset B),\Gamma\vdash M:\perp}{\D_1}}}}
\end{array}
$$
where $\Delta:=z:A\supset B,k':\neg B,\Gamma$.
\end{figure}
%-----------------------------------------------------

Suppose $\Es$ in the redex of $\rhos$ has (and the reduction happens at) type $\perp$: in this case, if $\rhos$ is applied, the inference $\Delta k'.-$ in the contractum concludes $\perp$, with $k:\neg\perp$. We separate rules $\rhos^{\perp}$ for this case, which avoid such unnecessary uses of RAA. An alternative would have been to adopt another reduction rule
$$
(\rho_3)\qquad\Delta k^{\neg\perp}.M\to[I/k]M\qquad\qquad(I:=\lb x.x)\enspace.
$$
Then, $\rhos^{\perp}$ would have become a derived rule
%$$
\begin{equation}\label{eq:derive-rho-3}
\begin{array}{rcl}
\Es^{\perp}\hole{\Delta k.M}&\to_{\rhos}&\Delta k'.[\lb z.k'(\Es^{\perp}\hole z)/k]M\\
&\to_{\rho_3}&[\lb z.I(\Es^{\perp}\hole z)/k]M\\
&\to_{\betai}^*&[\lb z.\Es^{\perp}\hole z/k]M
\end{array}
\end{equation}
%$$
Still, we do not adopt $\rho_3$ (in this, we agree with \cite{RehofSorensen94}). This is technically convenient, as we point out later.

As we leave out $\rho_3$, we might have considered forbidding the formation of $\Delta k.M$ with $k:\neg\perp$, and insisted that rules $\rhos$ should only happen at types different from $\perp$ - the latter seems even necessary to guarantee the confluence property. Since we only care about strong normalization, we also refrain from enforcing such restrictions.
%If the user of the system wants to stay in the fragment free from those instances of $\Delta k.M$, (s)he is very welcome: the special rules $\rhos^{\perp}$ ensure that the undesired expressions are never generated by reduction.

Later, we will consider another reduction relation, generated by the rule
$$
(\rho_4)\qquad k'(\Delta k.M)\to[k'/k]M\enspace.
$$
This will be just a technical device to help reasoning. This rule does not belong to the system.

If $R$ is a(n union of) reduction rule(s)\footnote{We will often
denote (long) unions by juxtaposition, \emph{e.g.} $R_1R_2R_3$
instead of $R_1\cup R_2\cup R_3$.}, then $\to_R$ denotes
\emph{$R$-reduction}, the binary relation inductively generated by
closing $R$ under all contexts. Then $\to_R^=$ (resp. $\to_R^+$,
$\to_R^*$) denotes the reflexive (resp. transitive, reflexive-transitive) closure of $\to_R$, whereas the latter's composition with itself $n-1$ times is denoted $\to_R^n$, as usual. $=_R$ is a
coarser notion of equality over proof expressions than ``syntactic''
identity. The later is denoted by $=$ (sometimes by $\equiv$ for
emphasis) and coincides with $\alpha$-equivalence, since we work
modulo the name of bound variables.

%----------------------------------------
\begin{defn}[Full, disjunction-free, and small systems]\label{def:systems}\quad
\begin{enumerate}
\item The system defined so far is denoted $\ld$: it is a presentation
of propositional classical natural deduction. We also refer to this system as the full system.

\item By $\ldmd$ we
denote the restriction of $\ld$ obtained by leaving out disjunction.
More precisely, in $\ldmd$ we omit: type $A\vee B$ and its term
constructions and typing rules; and the reduction rules $\betad$,
$\pis$, and $\Deltad$. In $\ldmd$, $\beta:=\betai\cup\betac$, $\rho_1:=\rhoi\cup\rhoc$, and $\rho_1^{\perp}:=\rhoi^{\perp}\cup\rhoc^{\perp}$. $\ldmd$ is also called the disjunction-free system.
%$$
%\begin{array}{rcl}
%\beta&:=&\betai\cup\betac\\
%\Delta&:=&\Delta_2\cup\Delta_3\\
%\eta&:=&\etal\cup\etad
%\end{array}
%$$
%The omission of $\Delta_1$ is going to be justified below.
\item The small system, denoted $\ldmdc$, is obtained from $\ldmd$ by leaving out conjunction, with implication and absurdity the logical constants remaining. In $\ldmdc$, $\beta:=\betai$, and $\rho_1:=\rhoi$. Furthermore, $\rhoi^{\perp}$ is omitted, hence there is no $\rho^{\perp}$ in this system.
\end{enumerate}
\end{defn}
%----------------------------------------

The full system will be the source of the De Morgan translation, whose target is the disjunction-free system. Strong normalization for our particular disjunction-free system is not a result found off the shelf - it will be proved as a corollary to strong normalization of the small system, the latter being a known result:% \cite{RehofSorensen94}:

%-------------------------------------
\begin{thm}\label{thm:SN-simple-system}
The small system $\ldmdc$ enjoys strong normalization.
\end{thm}
%-------------------------------------

\noindent\textbf{Proof.} $\ldmdc$ is system $\lb_{\Delta}^{\perp,\supset}$ of \cite{RehofSorensen94} minus its reduction rule (4), so Theorem 23 of \emph{op. cit.} applies. $\blacksquare$\\ 
\section{Lifting of strong normalization}\label{sec:lifting-of-SN}

%===============================================================
\subsection{The De Morgan translation}

We now present the translation into the disjunction-free sub-system. The \emph{de Morgan translation} consists of the obvious translation
of formulas
$$
\begin{array}{rcl}
\dmg X&=&X\\
\dmg \perp&=&\perp\\
\dmg{(A\supset B)}&=&\dmg A\supset\dmg B\\
\dmg{(A\wedge B)}&=&\dmg A\wedge\dmg B\\
\dmg{(A\vee B)}&=&\neg(\neg\dmg A\wedge\neg\dmg B)
\end{array}
$$

\noindent together with a translation of proof expressions given in
Fig.~\ref{fig:translation}. The translation is defined
homomorphically in all cases but those relative to the constructors
representing introduction and elimination of disjunction; and, for
these, the translation corresponds to the transformations recalled
in Fig.~\ref{fig:de-Morgan-map-of-proof-rules}.

%------------------------
\begin{figure}[t]\caption{The de Morgan translation of proof expressions}\label{fig:translation}
$$
\begin{array}{rcll}
\dmg x&=&x&\\
\dmg{(\lb x.M)}&=&\lb x.\dmg M&\\
\dmg{\pair MN}&=&\pair{\dmg M}{\dmg N}&\\
\dmg{(\inj iM)}&=&\lb w.\pi_i(w)\dmg M&\textrm{($i=1,2$)}\\
\dmg{(\Es\hole M)}&=&\dmg{\Es}\hole{\dmg M}&(\connective=\supset,\wedge)\\
\dmg{(\case MxPyQ)}&=&\left\{\begin{array}{l}\dmg M\pair{\lb x.\dmg P}{\lb y.\dmg Q}\\\Delta k.\dmg M\pair{\lb x.k\dmg P}{\lb y.k\dmg Q}\end{array}\right.&\begin{array}{ll}(P,Q:\perp)\\ \textrm{otherwise}\end{array}\\
%\dmg{(\case MxPyQ)}&=&\Delta k.\dmg M\pair{\lb x.k\dmg P}{\lb y.k\dmg Q}&\\
%\dmg{(\case MxPyQ)}&=&\dmg M\pair{\lb x.\dmg P}{\lb y.\dmg Q}&(P,Q:\perp)\\
\dmg{(\Delta k.M)}&=&\Delta k.\dmg M&\\
&&&\\
\dmg{(\ehole N)}&=&\ehole\dmg N\\
\dmg{(\pi_i(\ehole))}&=&\pi_i(\ehole)
\end{array}
$$
\end{figure}
%----------------------------------

%Typing helps one grasping what happens in de Morgan translation of
%proof expressions.

Notice that the target system of this translation is classical, therefore the translation is not a negative/CPS translation.

The case distinction in the definition of $\dmg{(\case MxPyQ)}$ means that we can only apply the translation of proof expressions to typed expressions, $M$ say, and in that case $\dmg M$ has the same type as $M$.

%------------------
\begin{prop}[Soundness]\label{prop:soundness}
The typing rules of Fig.~\ref{fig:typing-of-translation} are
derivable.
\end{prop}

%---------------------------------
\begin{figure}[t]\caption{Typing for the de Morgan translation}\label{fig:typing-of-translation}
$$
\begin{array}{c}
\infer{\dmg\Gamma\vdash\dmg M:\dmg A}{\Gamma\vdash
M:A}\qquad\infer[\connective=\supset,\wedge]{\dmg\Gamma|\dmg
A\vdash\Es:\dmg B}{\Gamma|A\vdash\Es:B}
\end{array}
$$
\end{figure}
%----------------------------------

The first of the typing rules in
Fig.~\ref{fig:typing-of-translation} states the logical soundness of
the translation: if $A$ is a theorem with proof $M$ in the source
system, then $\dmg A$ is a theorem with proof $\dmg M$ in the
disjunction-free system.

%--------
\begin{lemma}[Preservation of substitution]\label{lem:preservation-of-subst}$[\dmg N/x]\dmg M=\dmg{([N/x]M)}$.
\end{lemma}

\noindent\textbf{Proof.} We need in $\ld$ the obvious concept
$[Q/x]\E$:
$$
\begin{array}{rcl}
{[}Q/x](\ehole N)&=& \ehole[Q/x]N \\
{[}Q/x](\pi_i(\ehole))&=& \pi_i(\ehole) \\
{[}Q/x](\case{\ehole}{x_1}{P_1}{x_2}{P_2})&=& \case{\ehole}{x_1}{[Q/x]P_1}{x_2}{[Q/x]P_2}\\
\end{array}
$$
\noindent Then the following holds
$$
[Q/x](\E\hole M)=([Q/x]\E)\hole{[Q/x]M}\enspace.\qquad\qquad(*)
$$
Now, the lemma's statement is proved together with $[\dmg
N/x]\dmg{\Es}=\dmg{([N/x]\Es)}$, with $\connective=\supset,\wedge$,
by
simultaneous induction on $M$ and $\Es$. All cases follow by definitions, IHs, and fact $(*)$. $\blacksquare$\\

%----------
%\begin{figure}\caption{Correspondence of reduction rules for translation of proof reduction}\label{fig:simulation}
%$$
%\begin{array}{r|l}
%R & \dmg R\\
%\hline
%\betad & \beta\rho_2\\
%\pis & \betai\rhos\qquad(\connective=\supset,\wedge)
%\end{array}
%$$
%\end{figure}

%------------
\begin{thm}[Translation of proof reduction]\label{thm:translation-of-red}\quad
\begin{enumerate}
\item Let
$R\in\{\betai,\betac,\rhoi,\rhoi^{\perp},\rhoc,\rhoc^{\perp},\rho_2\}$ and
let $\dmg R=R$. If $M\to_R N$ in $\ld$, then $\dmg M\to_{\dmg R}\dmg
N$ in $\ldmd$.
\item Let
$R=\rhod^{\perp}$ and
let $\dmg R=\rhoi^{\perp}$. If $M\to_R N$ in $\ld$, then $\dmg M\to_{\dmg R}\dmg
N$ in $\ldmd$.
\item Let
$R=\betad$ (resp. $R=\pii$, $R=\pic$) and
let $\dmg R=\beta\rho_2$ (resp. $\dmg R=\betai\rhoi$, $\dmg R=\betai\rhoc$). If $M\to_R N$ in $\ld$, then $\dmg M\to_{\dmg R}^+\dmg
N$ in $\ldmd$.
%\item Let $R$ be one of the reduction rules in the left column of
%Fig.~\ref{fig:simulation}, and let $\dmg R$ be the corresponding
%union of reduction rules in the right column of the same figure. If
%$M\to_R N$ in $\ld$, then $\dmg M\to_{\dmg R}^+\dmg N$ in $\ldmd$.
\item Let $R\in\{\pid,\rhod\}$ and $\dmg R=\rhoi^{\perp}$. If $M\to_{R}N$ in $\ld$ then $\dmg M\to_{\dmg R}N'$ in $\ldmd$, for some $N'$ such that $\dmg N\to_{\rho_4}N'$.
\end{enumerate}
\end{thm}

\noindent\textbf{Proof.} By induction on $M\to_R N$. First we see the base cases.

Case $\betai$. Let $LHS:=(\lb x.M)N\to[N/x]M=:RHS$.
$$
\begin{array}{rcll}
\dmg{LHS}&=&(\lb x.\dmg M)\dmg N&\textrm{(by def.)}\\
&\to_{\betai}&[\dmg N/x]\dmg M&\\
&=&\dmg{([N/x]M)}&\textrm{(by Lemma \ref{lem:preservation-of-subst})}\\
&=&\dmg{RHS}&
\end{array}
$$

Case $\rhos$. Let $LHS:=\Es\hole{\Delta k.M}\to\Delta k'[\lb
z.k'(\Es\hole z)/k]M=:RHS$, with $\connective=\supset,\wedge$.
$$
\begin{array}{rcll}
\dmg{LHS}&=&\dmg{\Es}\hole{\Delta k.\dmg M}&\textrm{(by def.)}\\
&\to_{\rhos}&\Delta k'.[\lb z.k'(\dmg{\Es}\hole z)/k]\dmg M&\\
&=&\Delta k'.[\dmg{(\lb z.k'(\Es\hole z))}/k]\dmg M&\textrm{(by def.)}\\
&=&\Delta k'.\dmg{([(\lb z.k'(\Es\hole z))/k]M)}&\textrm{(by Lemma \ref{lem:preservation-of-subst})}\\
&=&\dmg{RHS}&\textrm{(by def.)}\\
\end{array}
$$

The other base cases of $R$ in statement 1 are equally straightforward.

Case $\rhod^{\perp}$. Let $LHS:=\case{\Delta k.M}xPyQ\to[\lb
z.\case zxPyQ/k]M=:RHS$. Let $N:=\pair{\lb x.\dmg P}{\lb
y.\dmg Q}$. Then:
$$
\begin{array}{rcll}
\dmg{LHS}&=&(\Delta k.\dmg M)N&\textrm{(by def.)}\\
%&\to_{\rhoi}&\Delta k'\Delta k_0.[\lb z.k_0(zN)/k]\dmg M&\\
%&\to_{\rho_3}&\Delta k'.[I/k_0][\lb z.k_0(zN)/k]\dmg M&\textrm{(since $k_0:\neg\perp$)}\\
%&=&\Delta k'.[\lb z.I(zN)/k]\dmg M&\\
%&\to_{\betai}^*&\Delta k'.[\lb z.zN/k]\dmg M&
&\to_{\rhoi^{\perp}}&[\lb z.zN/k]\dmg M&\\
&=&[\dmg{(\lb z.\case zxPyQ)}/k]\dmg M&\textrm{(by def.)}\\
&=&\dmg{RHS}&\textrm{(by Lemma \ref{lem:preservation-of-subst})}
\end{array}
$$

%Statement 2.

Case $\betad$. Let $LHS:=\case{\inj
iM}{x_1}{P_1}{x_2}{P_2}\to[M/x_i]P_i=:RHS$.
$$
\begin{array}{rcll}
\dmg{LHS}&=&\Delta k.(\lb w.\pi_i(w)\dmg M)\pair{\lb x_1.k\dmg{P_1}}{\lb x_2.k\dmg{P_2}}&\textrm{(by def.)}\\
&\to_{\beta}^3&\Delta k.k([\dmg M/x_i]\dmg{P_i})&\\
&\to_{\rho_2}&[\dmg M/x_i]\dmg{P_i}&\\
&=&\dmg{([M/x_i]P_i)}&\textrm{(by Lemma \ref{lem:preservation-of-subst})}\\
&=&\dmg{RHS}&\\
\end{array}
$$

Case $\pis$. Let $LHS:=\Es\hole{\case MxPyP}\to\case Mx{\Es\hole
P}y{\Es\hole Q}=:RHS$, with $\connective=\supset,\wedge$.
$$
\begin{array}{rcll}
\dmg{LHS}&=&\dmg{\Es}\hole{\Delta k.\dmg M\pair{\lb x.k\dmg P}{\lb y.k\dmg Q}}&\textrm{(by def.)}\\
&\to_{\rhos}&\Delta k'[\lb z.k'(\dmg \Es\hole z)/k](\dmg M\pair{\lb x.k\dmg P}{\lb y.k\dmg Q})&\\
&=&\Delta k'.\dmg M\pair{\lb x.(\lb z.k'(\dmg \Es\hole z))\dmg P}{\lb y.(\lb z.k'(\dmg \Es\hole z))\dmg Q}&\\
&\to_{\betai}^2&\Delta k'.\dmg M\pair{\lb x.k'(\dmg \Es\hole{\dmg P})}{\lb y.k'(\dmg \Es\hole{\dmg Q})}&\\
&=&\dmg{RHS}&\textrm{(by def.)}\\
\end{array}
$$

%Statement 3.

Case $\rhod$. Let $LHS:=\case{\Delta k.M}xPyQ\to\Delta k'[\lb
z.k'\case zxPyQ/k]M=:RHS$. Let $N:=\pair{\lb x.k'\dmg P}{\lb
y.k'\dmg Q}$. Then:
$$
\begin{array}{rcll}
\dmg{LHS}&=&\Delta k'.(\Delta k.\dmg M)N&\textrm{(by def.)}\\
%&\to_{\rhoi}&\Delta k'\Delta k_0.[\lb z.k_0(zN)/k]\dmg M&\\
%&\to_{\rho_3}&\Delta k'.[I/k_0][\lb z.k_0(zN)/k]\dmg M&\textrm{(since $k_0:\neg\perp$)}\\
%&=&\Delta k'.[\lb z.I(zN)/k]\dmg M&\\
%&\to_{\betai}^*&\Delta k'.[\lb z.zN/k]\dmg M&
&\to_{\rhoi^{\perp}}&\Delta k'.[\lb z.zN/k]\dmg M&
\end{array}
$$
\noindent On the other hand:
$$
\begin{array}{rcll}
\dmg{RHS}&=&\Delta k'.\dmg{([\lb z.k'\case zxPyQ/k]M)}&\textrm{(by def.)}\\
&=&\Delta k'.[\dmg{(\lb z.k'\case zxPyQ)}/k]\dmg M&\textrm{(by Lemma \ref{lem:preservation-of-subst})}\\
&=&\Delta k'.[\lb z.k'(\Delta k.z\pair{\lb x.k\dmg P}{\lb y.k\dmg Q})/k]\dmg M&\textrm{(by def.)}\\
&\to_{\rho_4}&\Delta k'.[\lb z.[k'/k]z\pair{\lb x.k\dmg P}{\lb y.k\dmg Q}/k]\dmg M&\\
&=&\Delta k'.[\lb z.zN/k]\dmg M&
\end{array}
$$

Case $\pid$ is proved by a similar argument.

As to inductive cases, it suffices to say that all the relations that hold between $\dmg M$ and $\dmg N$, namely
\begin{itemize}
\item $\cdot\to_R\cdot$
\item $\cdot\to_R^+\cdot$
\item $\exists N(\cdot\to_RN\wedge\cdot\to_{\rho_4}N)$
\end{itemize}
are congruences (i.e. compatible with the syntactic formation operations). So all the inductive cases follow routinely by induction hypothesis. $\blacksquare$\\

%Unfortunately, because of statement 4, the previous theorem does not
%yield a map of reduction sequences. However, the following
%is immediate.

%-------------
%\begin{cor}[Translation of proof identity]\label{cor:interpretation-of-identity} Let
%$R$ (resp. $\dmg{R}$) be the
%union of all reduction reduction rules of $\ld$ (resp. $\ldmd$). If
%$M=_{R}N$ in $\ld$, then $\dmg M=_{\dmg R}\dmg N$ in $\ldmd$.
%\end{cor}

%===============================================================
\subsection{Optimization}

Statement 4 of Theorem \ref{thm:translation-of-red} is an obstacle for the ready lifting of strong normalization. We now overcome this obstacle.

%----------------------------------------------
\begin{lemma}[Commutation of reduction steps]\label{lem:com} In $\ld$:
\begin{enumerate}
\item Let $R$ be a reduction rule different from $\rho_2$. If $M\to_{\rho_4}N_1$ and $M\to_RN_2$, then there is $N_3$ such that $N_1\to_RN_3$ and $N_2\to_{\rho_4}^*N_3$.
\item If $M\to_{\rho_4}N_1$ and $M\to_{\rho_2}N_2$, then there is $N_3$ such that: (i) $N_1\to_{\rho_2}N_3$ or $N_1=N_3$; and (ii) $N_2\to_{\rho_4}^*N_3$.
\end{enumerate}
\end{lemma}
%----------------------------------------------

\noindent\textbf{Proof.} There is only one case where a $\rho_4$-redex overlaps non-trivially with another $R$-redex, which is when $R=\rho_2$ and $M=k'(\Delta k.kM)$, with $k\notin M$. In this case, take $N_3=k'M$. $\blacksquare$\\

If $s$ is a reduction sequence, let $|s|$ denote its length, that is the number of reduction steps in $s$.

%----------------------------------------------
\begin{thm}[Translation of reduction sequences]\label{thm:length-preserving}
If $s$ is a reduction sequence in $\ld$ from $M$ to $N$, then there is $N'$ such that:
\begin{enumerate}
\item There is a reduction sequence $s'$ in $\ldmd$ from $\dmg M$ to $N'$, and $|s'|\geq|s|-m$, where $m$ is the number of $\rho_2$-reduction steps in $s$.
\item $\dmg N\to_{\rho_4}^*N'$.
\end{enumerate}
\end{thm}
%----------------------------------------------
\noindent\textbf{Proof.} By induction on $|s|$. The base case $|s|=0$ is trivial, just take $N'=\dmg N$. The inductive case follows from this diagram (where double-headed arrows denote $\to^*$):
%\newarrow{Onto}----{>>}
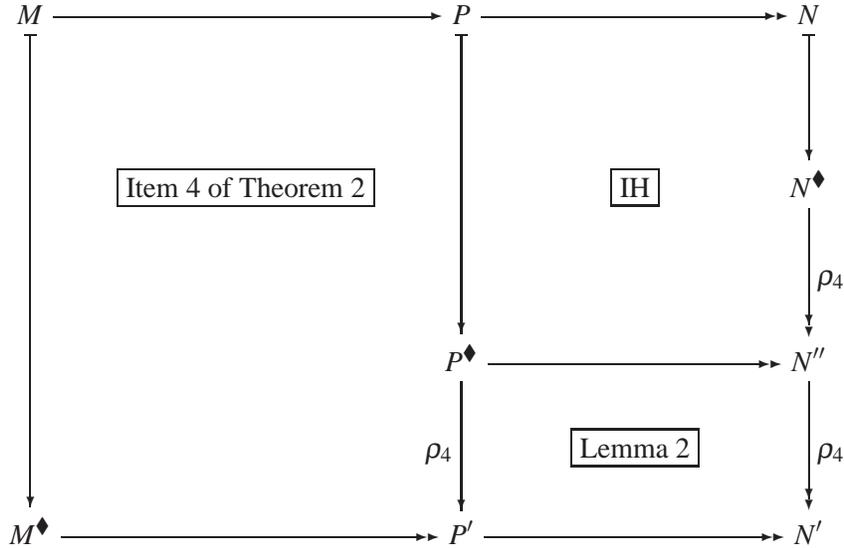
\begin{diagram}
M&&\rTo&&P&&\rOnto&&N\\
&&&&&&&&\dMapsto\\
&&\fbox{Item 4 of Theorem \ref{thm:translation-of-red}}&&\dMapsto&&\fbox{IH}&&\dmg N\\
&&&&&&&&\dOnto_{\rho_4}\\
\dMapsto&&&&\dmg P&&\rOnto&&N''\\
&&&&\dTo^{\rho_4}&&\fbox{Lemma \ref{lem:com}}&&\dOnto_{\rho_4}\\
\dmg M&&\rOnto&&P'&&\rOnto&&N'\\
\end{diagram}
If any other item of Theorem \ref{thm:translation-of-red} applies instead, then $\dmg M\to^+\dmg P$, and Lemma \ref{lem:com} is not needed. $\blacksquare$\\

This theorem says that a reduction sequence in the full system from a proof $M$ determines, essentially in a length-preserving way, a reduction sequence in the disjunction-free system from the De Morgan translation $\dmg M$ of the proof.

%----------------------------------------------
\begin{cor}[Lifting of strong normalization]\label{cor:lifting}
If the disjunction-free system enjoys strong normalization, so does the full system.
\end{cor}
\noindent\textbf{Proof.} Suppose there is an infinite reduction sequence from typable $M$ in the full system $\ld$. Thanks to Proposition \ref{prop:soundness}, $\dmg{(\cdot)}$ preserves typability, so $\dmg M$ is typable. We prove that, for any $n$, there is in $\ldmd$ a reduction sequence $s'$ from $\dmg M$ of length $n$. The existence of an infinite reduction sequence from $\dmg M$ then follows by K\"onig's Lemma. Let $n$ be given. Then, there is an initial segment $s$ of the infinite reduction sequence from $M$ such that $|s|-m\geq n$, where $m$ is the number of $\rho_2$-reduction steps in $s$. That such $s$ exists follows from termination of $\rho_2$-reduction. From the previous theorem, there is in $\ldmd$ a reduction sequence $s'$ from $\dmg M$ such that $|s'|\geq|s|-m$. $\blacksquare$\\

This is the positive answer to the main question raised in this paper.

\section{Strong normalization}\label{sec:SN}

We confirm that the target of the De Morgan translation is a system enjoying strong normalization.

Since we want to lift strong normalization from $\ldmdc$ (Theorem \ref{thm:SN-simple-system}), we have to translate $\ldmd$ into $\ldmdc$, i.e., we have to get rid of conjunction. Following \cite{RehofSorensen94}, we use the map $\mc{(\cdot)}$ of formulas induced by
$$
\mc{(A\wedge B)}=\neg(\mc A\supset\neg\mc B)\enspace.
$$

The map of proofs is defined by
$$
\begin{array}{rcll}
\mc{\pair MN}&=&\lb f.f\mc M\mc N\\
\mc{\pi_i(M)}&=&\Delta k.\mc M(\lb x_1x_2.kx_i)\\
\mc{\pi_i(M)}&=&\mc M(\lb x_1x_2.x_i)&(M:A_1\wedge A_2,\quad A_i=\perp)
\end{array}
$$
with the remaining cases defined homomorphicaly. It turns out that the study of this map follows the same patterns as that of $\dmg{(\cdot)}$. We will be very brief now.

%------------
\begin{thm}[Translation of proof reduction]\label{thm:translation-of-red-2} Let $S:=\ldmdc + \rhoi^{\perp}$.
\begin{enumerate}
\item Let $R$ be a reduction rule of $\ldmd$ different from $\rhoc$. If $M\to_R N$ in $\ldmd$, then $\mc M\to^+\mc N$ in $S$.
\item If $M\to_{\rhoc}N$ in $\ldmd$ then $\mc M\to_{\rhoi^{\perp}}N'$ in $S$, for some $N'$ such that $\mc N\to_{\rho_4}N'$.
\end{enumerate}
\end{thm}

\noindent\textbf{Proof.} By induction on $M\to_RN$. $\blacksquare$\\

Why do we temporarily need reduction rule $\rhoi^{\perp}$ in the target of $\mc{(\cdot)}$? Not only because the rule exists in the source calculus $\ldmd$, but also because it is needed to map $\rhoc$ and $\rhoc^{\perp}$, in the same way as it happened before with $\dmg{(\cdot)}$ and rules $\rhod$ and $\rhod^{\perp}$ (recall items 2 and 4 of Theorem \ref{thm:translation-of-red}).

Before, we have used Lemma \ref{lem:com} to obtain Theorem \ref{thm:length-preserving} from Theorem \ref{thm:translation-of-red}. Exactly in the same way, we use Lemma \ref{lem:com} to obtain from Theorem \ref{thm:translation-of-red-2} the following: if $s$ is a reduction sequence from $M$ in $\ldmd$, then there is a reduction sequence $s'$ in $\ldmdc + \rhoi^{\perp}$ from $\mc M$ with length $|s'|\geq|s|-m$, where $m$ is the number of $\rho_2$-reduction steps in $s$. Now, $\rhoi^{\perp}$ becomes a derived rule, if we accept (temporarily) rule $\rho_3$ (recall (\ref{eq:derive-rho-3})). Hence, we can rephrase the result just obtained:

%----------------------------------------------
\begin{thm}[Translation of reduction sequences]\label{thm:length-preserving-2} Let $S:=\ldmdc + \rho_3$. If $s$ is a reduction sequence from $M$ in $\ldmd$, then there is a reduction sequence $s'$ in $S$ from $\mc M$ with length $|s'|\geq|s|-m$, where $m$ is the number of $\rho_2$-reduction steps in $s$.
\end{thm}
%----------------------------------------------

Rule $\rho_3$ makes a timely appearance, now that the commutation argument of Lemma \ref{lem:com} is no longer needed. If the rule were present from the beginning, the commutation of $\rho_3$ and $\rho_4$ would fail, and the on-going strategy would be wrong.

%----------------------------------------------
\begin{cor}[Lifting of strong normalization]\label{cor:lifting-2}
If $\ldmdc+\rho_3$ enjoys strong normalization, so does $\ldmd$.
\end{cor}
\noindent\textbf{Proof.} From the previous theorem, termination of $\rho_2$-reduction, and preservation of typability by $\mc{(\cdot)}$ . $\blacksquare$\\

Since strong normalization of $\ldmdc$ (Theorem \ref{thm:SN-simple-system}) does not comprehend $\rho_3$, the last task is to get rid of this rule. This is done by postponement.

%-------------------
\begin{lemma}[Postponement]\label{lem:post} In $\ldmdc + \rho_3$, let $\kappa$ be the rule $kN\to N$ at type $\perp$.
\begin{enumerate}
\item If $M\to_{\rho_3}P\to_RQ$, then there is $P'$ such that $M\to_{R'}P'\to_{\rho_3}^*Q$, where $R'=R$, except in the case $R=\betai$, of the particular form $IN\to N$ at type $\perp$, for which $R'$ may be $R$ or $\kappa$.
\item If $M\to_{\kappa}P\to_RQ$, then there is $P'$ such that $M\to_{R}P'\to_{\kappa}^*Q$
\end{enumerate}
\end{lemma}

\noindent\textbf{Proof.} The exception mentioned in item 1 occurs when $M=\Delta k^{\neg\perp}.M'$ and $P=[I/k]M'$, with $Q$ resulting from the reduction of some $IN$ created by substitution $[I/k]\_$. Let $M''$ be the result of applying in $M'$, to the relevant occurrence of $k$, the rule $kN\to N$. Put $P'=\Delta k.M''$. $\blacksquare$\\

%----------------------------------------------
\begin{cor}[Lifting of strong normalization]\label{cor:lifting-3}
If $\ldmdc$ enjoys strong normalization, so does $\ldmdc + \rho_3$.
\end{cor}
\noindent\textbf{Proof.} Suppose there is an infinite reduction sequence from $M$ in $\ldmdc + \rho_3$. We prove that, for any $n$, there is in $\ldmdc$ a reduction sequence $s'$ from $M$ of length $n$. Let $n$ be given. Let $\iota$ be the rule $IN\to N$ at type $\perp$. There is an initial segment $s$ of the infinite reduction sequence from $M$ such that $|s|-m-i\geq n$, where $m$ is the number of $\rho_3$-reduction steps in $s$ and $i$ is the number of $\iota$-reduction steps in $s$. That such $s$ exists follows from termination of $(\rho_3\cup\iota)$-reduction. By applying postponement of $\rho_3$ (item 1 of Lemma \ref{lem:post}) to $s$, one obtains a reduction sequence $s'$ from $M$, without $\rho_3$-steps, of length $|s|-m$, where some $\iota$-steps are converted into $\kappa$-steps. The number of steps in $s'$ which are not $\kappa$-steps is a number $n'\geq|s|-m-i$. By postponement of $\kappa$ (item 2 of Lemma \ref{lem:post}) applied to $s$, we get a reduction sequence $s''$ from $M$, without $\kappa$-steps, of length $n'$. This is a reduction in $\ldmdc$ with length $\geq n$. $\blacksquare$\\

%-------------------------------------
\begin{thm}\label{thm:SN-disjunction-free}
The disjunction-free system $\ldmd$ enjoys strong normalization.
\end{thm}

\noindent\textbf{Proof.} From Corollaries \ref{cor:lifting-2} and \ref{cor:lifting-3} and Theorem \ref{thm:SN-simple-system}. $\blacksquare$\\

%-------------------------------------
\begin{thm}The full system $\ld$ enjoys strong normalization.
\end{thm}

\noindent\textbf{Proof.} From Corollary \ref{cor:lifting} and Theorem \ref{thm:SN-disjunction-free}. $\blacksquare$\\ 
%\input{first-variant}
%\input{second-variant}
%\input{conclusions}

%\appendix
%\input{appendix}

%\providecommand{\urlalt}[2]{\href{#1}{#2}}
%\providecommand{\doi}[1]{doi:\urlalt{http://dx.doi.org/#1}{#1}}

\textbf{Acknowledgments:} This research was financed by Portuguese Funds through FCT — Funda\c c\~ao para a Ci\^encia e a Tecnologia, within the Project UID/MAT/00013/2013.

\bibliography{SN-CL}
\bibliographystyle{eptcs}

\end{document}